\documentclass[10pt,letterpaper]{article}
\usepackage[top=0.85in,left=0.75in,footskip=0.75in,marginparwidth=0.5in]{geometry}

\usepackage[utf8]{inputenc}
\usepackage{amsmath,amssymb}

\usepackage{cite}

\usepackage{microtype}
\DisableLigatures[f]{encoding = *, family = * }

\raggedright
\usepackage{changepage}

\usepackage[aboveskip=1pt,labelfont=bf,labelsep=period,singlelinecheck=off]{caption}

\makeatletter
\renewcommand{\@biblabel}[1]{\quad#1.}
\makeatother

\usepackage{lastpage,fancyhdr,graphicx}
\usepackage{epstopdf}
\usepackage{color}

\definecolor{Gray}{gray}{.25}

\usepackage{graphicx}

\usepackage{sidecap}

\usepackage{wrapfig}
\usepackage[pscoord]{eso-pic}
\usepackage[fulladjust]{marginnote}
\reversemarginpar

\begin{document}
\vspace*{0.35in}

% title goes here:
\begin{flushleft}
{\Large
\textbf\newline{Multiple-timescale Neural Networks: Generation of Context-dependent Sequences and Inference through
Autonomous Bifurcations}
}
\newline
% authors go here:
\\
Tomoki Kurikawa\textsuperscript{1,*},
Kunihiko Kaneko\textsuperscript{2,3}
\\

\bigskip
\bf{1} Department of Physics, Kansai Medical University, Shinmachi 2-5-1, Hirakata, Osaka, Japan
\\
\bf{2} Department of Basic science, Graduate school of arts and sciences, University of Tokyo, Komaba 3-8-1, Meguro-ku, Tokyo, Japan
\\
\bf{3} Center for Complex Systems Biology, Universal Biology Institute, 
 University of Tokyo, Komaba 3-8-1, Meguro-ku, Tokyo, Japan

\bigskip
* kurikawa@hirakata.kmu.ac.jp

\end{flushleft}

\section*{Abstract}
Sequential transitions between metastable states are ubiquitously observed in the neural system and underlie various cognitive functions.
Although a number of studies with asymmetric Hebbian connectivity have investigated how such sequences are generated, the focused sequences are simple Markov ones.
On the other hand, supervised machine learning methods can generate complex non-Markov sequences, but these sequences are vulnerable against perturbations. Further, concatenation of newly learned sequence to the already learned one is difficult due to catastrophe forgetting, although concatenation is essential for cognitive functions such as inference.
How stable complex sequences are generated still remains unclear.
We have developed a neural network with fast and slow dynamics, which are inspired by the experiments.
The slow dynamics store history of inputs and outputs and affect the fast dynamics depending on the stored history.
We show the learning rule that requires only local information can form the network generating the complex and robust sequences in the fast dynamics.
The slow dynamics work as bifurcation parameters for the fast one, wherein they stabilize the next pattern of the sequence before the current pattern is destabilized.
This co-existence period leads to the stable transition between the current and the next pattern in the sequence.
We further find that timescale balance is critical to this period.
Our study provides a novel mechanism generating the robust complex sequences with multiple timescales in neural dynamics.
Considering the multiple timescales are widely observed, the mechanism advances our understanding of temporal processing in the neural system.

% now start line numbers
%\linenumbers

% the * after section prevents numbering
\section*{Introduction}
Sequentially activated patterns are widely observed in neural systems, for instance, cerebral cortex \cite{Jones2007,Mazzucato2015,Ponce-Alvarez2012,Kurikawa2018,Taghia2018,Stokes2013}, hippocampus \cite{Maboudi2018,Gupta2010,Wimmer2020,Schuck2019} and the striatum \cite{Akhlaghpour2016}.
These patterns underlie a range of cognitive functions:
perception \cite{Jones2007,Miller2010}, decision making \cite{Ponce-Alvarez2012}, working memory \cite{Stokes2013,Taghia2018}, and recall of long-term memory \cite{Wimmer2020}.
They process temporal information by concatenating shorter sequences \cite{Gupta2010}, reorganizing the order in sequential patterns \cite{Wimmer2020}, and chunking sequences \cite{Jin2014}, which lead to inference and recall based on previous experience.

Several models have been proposed to understand how such sequential patterns are shaped in the neural systems in order to perform complex tasks \cite{Kleinfeld1986, Sompolinsky1986, Seliger2003, Recanatesi2015, Russo2012, Haga2019, Gros2007,Sussillo2009,Laje2013,Chaisangmongkon2017}.
Popular Hebbian models provide a simple framework in which each pattern in sequence is represented as a metastable state, which is formed through Hebbian learning.
An asymmetric connection from the current to the successive pattern \cite{Amari1972,Kleinfeld1986, Sompolinsky1986, Nishimori1990, Seliger2003, Recanatesi2015, Russo2012, Haga2019, Gros2007} causes transition between patterns, as well as adaptation terms \cite{Gros2007, Recanatesi2015, Russo2012} and the winnerless competition \cite{Seliger2003}.
These sequences are robust to noise and widely observed in the neural systems \cite{Miller2016}.
In other studies \cite{Mante2013,Chaisangmongkon2017,Sussillo2009,Laje2013}, recurrent neural networks (RNN) are trained using machine learning methods to generate neural trajectories.
RNNs reproduce neural behaviors measured in cortical areas, and how neural trajectories encode and process information in time has been investigated.

In spite of great success of these studies, however, some fundamental questions remain unanswered.
In models that generate sequential metastable states, a transition between these states is embedded rigidly into the connectivity (i.e., correlation between the current to the next pattern), resulting in 
successive patterns being determined by the immediately preceding pattern.
Hence, generation of sequences depending on the long history of the previous patterns is not possible.
On the other hand, RNNs trained using machine learning methods allow for generating  complex sequences dependent on the history.
In these RNNs, however, the previous learned patterns are easily erased upon learning new patterns. 
Thus, connecting new sequences with the learned sequences as is necessary for inference is quite difficult.
Further, the training methods require non-local information, which is not biologically plausible,
and the formed sequences are vulnerable to noise or perturbation to the initial state \cite{Laje2013}.

\begin{figure*}[t]
\centering
\includegraphics[width=120mm]{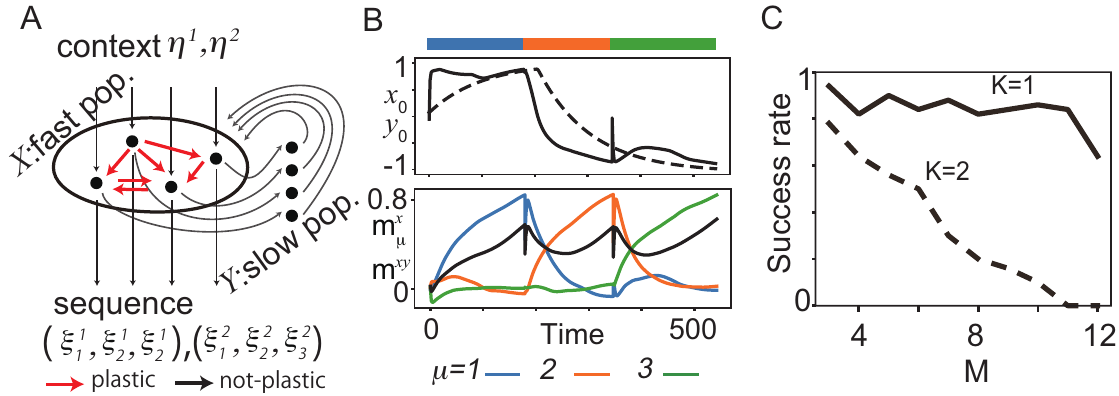}
\caption{
    {\textbf A}: Schematic diagram of the proposed model for $K=2, M=3$.
    {\textbf B}: Neural dynamics during the learning process of three targets.
    Top: the time series of one of the fast variables $x_{0}$ (solid line) and the corresponding slow variable $y_{0}$ (broken line) during the learning process.
    Bottom: $m_{0,1,2}^{x}$, overlaps of $\boldsymbol{x}$ with $\boldsymbol{\xi}^0_{0}$ (blue), $\boldsymbol{\xi}^0_{1}$ (orange), and $\boldsymbol{\xi}^0_{2}$ (green).
    The black line represents the overlap between $\boldsymbol{x}$ and $\boldsymbol{y}$ denoted as $m^{xy}$.
    The bars above the panels indicate the targeted patterns given to the network in corresponding periods.
    {\textbf C}:The fraction of successful recalls is plotted as a function of $M$ for $K=1,2$.
	It is averaged over 50 realizations (10 networks and five pairs of the target and context signal patterns for each network).
    Here, a successful recall is defined as the case in which all $K \times M$ targets are sequentially generated in the correct order in the presence of the corresponding context signals.
    }
    \label{fig:neural_beh}
\end{figure*}

To address this issue, we introduce a neural network model with slow and fast neurons, which can generate complex sequences robustly; this model is inspired by observations that the timescales in the neural activity change are distributed across cortical areas \cite{Chaudhuri2015, Hasson2015, Murray2014, Honey2012}.
The fast neural dynamics generate patterns in response to an external input.
The slow dynamics store the history of the inputs from the fast dynamics, and feed the stored information back to the fast, as shown in Fig. \ref{fig:neural_beh}A.
In fact, such multiple-timescale dynamics are observed across cortical areas \cite{Chaudhuri2015, Hasson2015, Murray2014, Honey2012}.
Neural activities in sensory cortices change in a faster timescale and respond instantaneously to stimuli,
whereas those in association cortices change in a slower timescale and integrate information over longer periods.
Although existence of multiple timescales is widely observed, its relevance to temporal information processing has not yet been fully explored.
Our study explores how the slow dynamics control the fast dynamics to generate complex sequences.

In particular, we focus on two basic aspects of neural sequences in temporal information processing.
First, generating context-dependent sequences is investigated.
In fast, neural systems respond differently to the same stimuli depending on the contexts.
In the context-dependent working memory task \cite{Mante2013,Stokes2013}, distinct sequences of neural patterns are evoked by identical stimuli depending on the preceding context signals.
We demonstrated how such a sequence with non-Markov property is generated.
Second, we investigate inference, the ability to make appropriate responses against new environment by using previously learned examples.
For instance \cite{Jones2012,Wikenheiser2016}, consider a rat learning successive stimuli, A followed by B, and then reward C.
After changing the environment, the rat is required to learn a new combination of stimuli, A' followed by B.
In this situation, the rat is able to infer that stimuli A' causes the reward C via B.
Neural activities reflecting this cognitive function should show sequential patterns A'BC even after learning only A'B.

\begin{figure*}[t]
\centering
\includegraphics{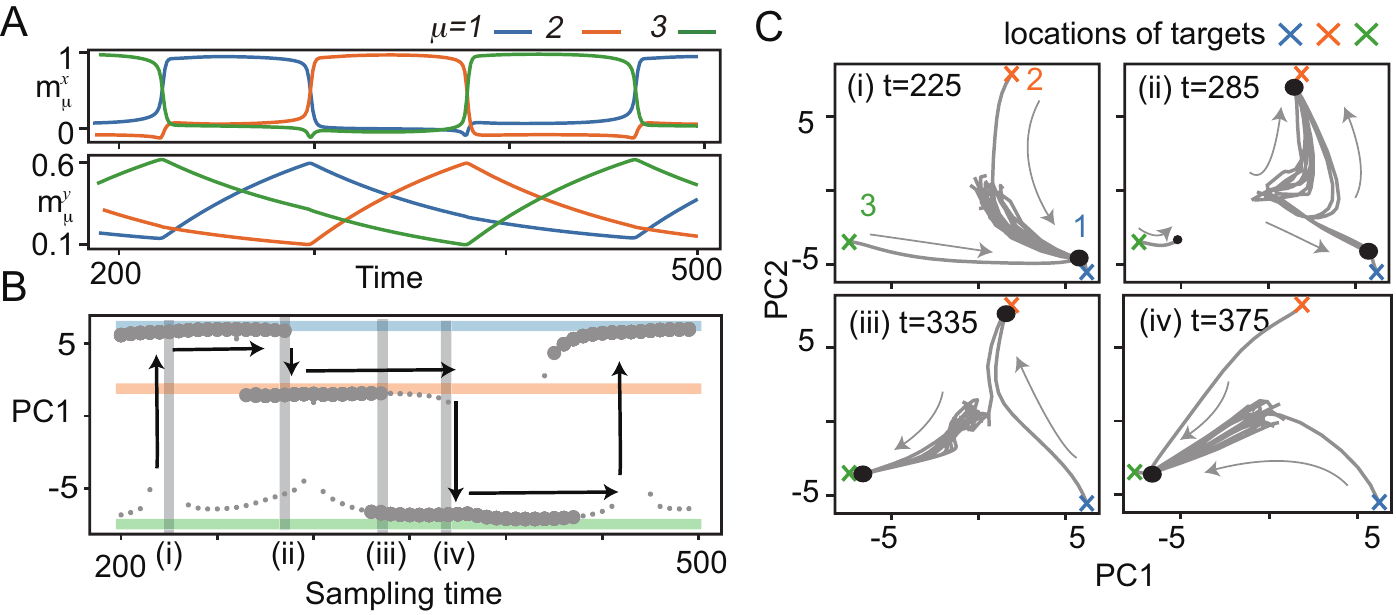}
    \caption{Bifurcation of $\boldsymbol{x}$ with quenched $\boldsymbol{y}$.
        {\textbf A}: Neural dynamics during the recall process of the three learned patterns.
        Overlaps of neural activities $m_{\mu}^{x,y}$, $\mu=1,2,3$ in $\boldsymbol{x}$ (top) and $\boldsymbol{y}$ (bottom) for $M=3$ are plotted in the same color as shown in Fig. \ref{fig:neural_beh}B.
        $\boldsymbol{y}$ is sampled from the trajectory at $200<t<500$ for the bifurcation diagram of $\boldsymbol{x}$ shown in B.
        {\textbf B}: Bifurcation diagram of $\boldsymbol{x}$ as quenched $\boldsymbol{y}$ is updated with the sampling time. 
        Fixed points of $\boldsymbol{x}$ are shown by projecting to the first principal component (PC1) of principle component analysis.
        Small circles indicate fixed points with small basins: neural activity beginning only from the vicinity of the target converges to these points.
        Large circles represent fixed points with large basins: neural activities from the initial states converge to these points.
        To identify fixed points, the neural states are plotted after the transient period.
        Colored lines indicate the locations of the targets ($\boldsymbol{\xi}_{1,2,3}^1$ in blue, orange, and green, respectively).
        Vertical arrows show the transitions of $\boldsymbol{x}$ to different targets in the recall process.
        {\textbf C}: 
        The neural dynamics for given a $\boldsymbol{y}$ at $t=225,285,335,375$ shadowed in B are depicted by projecting $\boldsymbol{x}$ to the 2-dimensional PC space (PC1 is same as that in B).
        Fifteen trajectories (three from the vicinity of the target, and others from random initial states) are plotted.
        Large and small circles represent fixed points given in B.
	}
  \label{fig:bif}
\end{figure*}

We study a multiple-timescale network that can learn the context-dependent sequences and connect the sequences, in which the slow dynamics control successive bifurcations of fixed points of fast dynamics, based on the stored history of previous patterns and inputs.
By adopting a biologically plausible learning rule based solely on the correlation between the pre- and post-synaptic neural activities introduced previously \cite{Kurikawa2013, Kurikawa2016, Kurikawa2020}, we demonstrate that our model with the fast and slow neural dynamics memorizes the history-dependent sequences and enables inference based on them.

\section*{Results}
We consider learning of $K$ sequences, each of which contains $M$ patterns, with $K$ context signals.
We denote the $\mu$-th targeted pattern in the $\alpha$-th sequence as $\boldsymbol{\xi}^{\alpha}_{\mu}$, and the corresponding context signal as $\boldsymbol{\eta}^{\alpha}$ for $\mu=1,2,\cdots, M$ over the inputs $\alpha=1, \cdots, K$.
Fig. \ref{fig:neural_beh}A illustrates the case with $K=2$ and $M=3$:
In this case, a given sequence $\boldsymbol{\xi}^{\alpha}_{1}$,$\boldsymbol{\xi}^{\alpha}_{2}$,$\boldsymbol{\xi}^{\alpha}_{3}$ ($\alpha=1,2$) should be generated upon a given corresponding context signal $\boldsymbol{\eta}^{\alpha}$.
Generally, a pattern to be generated next is determined not only by the current pattern, but also by earlier patterns.
Thus, a network has to retain the history of previous patterns to generate a sequence correctly.

To achieve this, we built a two-population model with different timescales, one with $N$ fast neurons and one with $N$ slow neurons, denoted as $X$ and $Y$, respectively.
$X$ receives an external input, and $Y$ receives the output from $X$ and provides input to $X$, as shown in Fig \ref{fig:neural_beh}A.
The neural activities $x_i$ in $X$ and $y_i$ in $Y$ evolve according to the following equation:
\begin{align}
  \tau_{x}\dot{x_{i}} &=& \tanh{(\beta_x I_{i})} - x_{i},  \label{eq:neuro-dyn}\\ 
  \tau_{y}\dot{y_{i}} &=& \tanh(\beta_y x_{i}) - y_{i},  \label{eq:neuro-dyn-slow}\\ 
  I_i  &=& u_i  +  \tanh(r_i) +(\eta^{\alpha})_{i},  \label{eq:input}
\end{align}
where $u_{i} = \sum_{j \neq i}^{N} J_{ij}^{X} x_{j}$;
$r_{i} = \sum_{j}^{N} J_{ij}^{XY} \tanh(y_{j})$
\footnote{For the input from $Y$ to $X$, we considered two nonlinear filters by the hyperbolic tangent function under the following biological assumptions.
First, for $\tanh(y_j)$, the activity of $y_i$ is assumed to be amplified in a nonlinear way at a synapse onto $x_i$.
Second, for $\tanh(r_i)$, we considered a large dendritic branch of $x_i$ to which all inputs from $Y$ are injected and assumed that activity of the branch (i.e., summation of total inputs from $Y$) surges beyond the threshold.}.
$J_{ij}^{X}$ is a recurrent connection from the $j$-th to the $i$-th neuron in $X$, and $J^{XY}_{ij}$ is a connection from the $i$-th neuron in $Y$ to the $j$-th neuron in $X$.
The mean values of $J^{X}$ and $J^{XY}$ are set at zero with the variance equal to $1/N$.
$X$ is required to generate the pattern $\boldsymbol{\xi}^{\alpha}_{\mu}$ in the presence of $\boldsymbol{\eta}^{\alpha}$,
i.e., an attractor that matches $\boldsymbol{\xi}^{\alpha}_{\mu}$ is generated. 
The $i$-th element of a targeted pattern, denoted as $(\xi^{\alpha}_{\mu})_i$, is assigned to the $i$-th neuron in $X$, and randomly sampled according to the probability $P[ (\xi^{\alpha}_{\mu})_{i}=\pm 1]=1/2$.
The context signal $(\eta^{\alpha})_i$ is injected to the $i$-th neuron in $X$, randomly sampled according to $P[(\eta^{\alpha})_i=\pm 1]=1/2$. 
We set $N=100, \beta_x=2, \beta_y=20, \tau_x=1,$ and $\tau_y=100$.

Only $J^{X}$ changes to generate the target according to the following equation: 
\begin{align}
	\tau_{syn}\dot{J_{ij}^{X}} &=& (1/N)(\xi_i - x_i)(x_j - u_i J_{ij}^X),
\end{align}
where $\tau_{syn}$ is the learning speed (set at $100$).
This learning rule comprises a combination of a Hebbian term between the target and the presynaptic neuron, and an anti-Hebbian term between the pre- and post-synaptic neurons with a decay term $u_i J_{ij}^X$ for normalization.
This form satisfies locality across connections, and is biologically plausible \cite{Kurikawa2020}.
We previously applied this learning rule to a single network of $X$, and demonstrated that the network learns $K$ maps between inputs and targets, i.e., $M=1$ \cite{Kurikawa2013,Kurikawa2016,Kurikawa2020}.
However, in that case, generating a sequence ($M \geq 2$) was not possible.
In the present study, there are two inputs for $X$, 
one from a context signal $\boldsymbol{\eta}$ and one from $Y$ that stores previous information.
Thus, the network can generate a pattern depending not only on the present input (context) signal, but also on the previous patterns.

Before exploring the history-dependent sequence, we analyzed if our learning rule generates simple sequences, namely, sequences in which the successive pattern is determined solely by the current pattern.
Fig \ref{fig:neural_beh}B shows a sample learning process for $K=1$.
We applied $\boldsymbol{\eta}^1$ to a network, and presented $\boldsymbol{\xi}^{1}_{1}$ as the first pattern of a target sequence.
After the transient time, $\boldsymbol{x}$ converges to $\boldsymbol{\xi}^{1}_{1}$ due to synaptic change.
$\boldsymbol{y}$ follows $\boldsymbol{x}$ according to Eq. \ref{eq:neuro-dyn-slow} and, consequently, moves to the target.
A learning step of a single pattern is accomplished when the neural dynamics satisfy the following two criteria:
$\boldsymbol{x}$ sufficiently approaches the target pattern,
i.e., $m_{\mu}^x \equiv \Sigma_i x_i (\xi^1_{\mu})_i/N > 0.85$,
and $\boldsymbol{y}$ is sufficiently close to $\boldsymbol{x}$, 
i.e., $\Sigma_i x_i y_i /N > 0.5$.
After the completion of one learning step, a new pattern $\boldsymbol{\xi}^1_2$ is presented instead of $\boldsymbol{\xi}^1_1$ with a perturbation of fast variables $x_i$, by multiplying a random number uniformly sampled from zero to one.
We execute these steps sequentially from $\mu=1$ to $M$ to learn a sequence.
The fast and slow variables learn the next sequence until this procedure is repeated 20 times for each sequence.

We, then, present an example of a recall process after the learning process for $(K,M)=(1,3)$ in Fig. \ref{fig:bif}A.
In recall, the connectivity is not changed.
The initial states of the fast variables are set at random values sampled from a uniform distribution of 0 to 1, whereas the slow variables are set at values of the final state of the learning process.
The targets appear sequentially in $X$ in order.
Note that in the recall process, the transition occurs spontaneously without any external operation.
The proposed model is able to memorize multiple sequences, say $M=11$ for $K=1$, and $M=3$ for $K=2$.

We explored the success rate of the learning, and found that increasing $M$ and $K$ generally leads to a decrease in the success rate of recalls.
For $N=100$ and $K=1$ up to $M=11$, the success rate is over 80\%, and decreases beyond $K=12$.
For $K=2$, the success rate is approximately 80\% for $M=3$, and decreases gradually as $M$ increases (Fig. \ref{fig:neural_beh}C, see the Supplemental Material for detailed results).
Furthermore, we investigated how the balance between the timescales of the slow variables $\tau_y$ and learning $\tau_{syn}$ affect the success rate.

To examine the robustness of the recall, we explored trajectories from different initial conditions with Gaussian white noise with strength $s$ (See Supplemental Material for details).
All of these trajectories converge correctly to a target sequence after some transient period for weak noise.
By increasing the noise strength, the recall performance of noisy dynamics is made equal to that of the noiseless dynamics up to noise strength $s=0.3$. Even upon applying strong and instantaneous perturbation to both $\boldsymbol{x}$ and $\boldsymbol{y}$, the trajectory recovers the correct sequence.
The sequence is represented as a limit cycle containing $\boldsymbol{x}$ and $\boldsymbol{y}$, and thus, is recalled robustly. 

\subsection*{Bifurcations of fast neural dynamics}
To elucidate how such a robust recall is possible,
we analyzed the phase space of $\boldsymbol{x}$ with $\boldsymbol{y}$ quenched.
In other words, $\boldsymbol{y}$ is regarded as bifurcation parameters for the fast dynamics.
Specifically, we focused on the neural dynamics for $200 \leq t \leq 500$, as shown in Fig. \ref{fig:bif}A.
In this period, the fast dynamics show transitions from $\boldsymbol{\xi}^1_1$ to $\boldsymbol{\xi}^1_2$ at $t=290$, from $\boldsymbol{\xi}^1_2$ to $\boldsymbol{\xi}^1_3$ at $t=375$, and from $\boldsymbol{\xi}^1_3$ to $\boldsymbol{\xi}^1_1$ at $t=220,460$.
We sampled the slow variables every five units of time from $t=200$ to $500$, $\boldsymbol{y}_{t=200},\boldsymbol{y}_{t=205}, \cdots, \boldsymbol{y}_{t=500}$,
along the trajectory, 
and analyzed the dynamics of $\boldsymbol{x}$ with the slow variables quenched at each sampled $\boldsymbol{y}_{t=200,205, \cdots, 500}$.
Fig. \ref{fig:bif}B shows the bifurcation diagram of $\boldsymbol{x}$ against the change in $\boldsymbol{y}$, and Fig. \ref{fig:bif}C shows the the trajectories of $\boldsymbol{x}$ for specific $\boldsymbol{y}$.

\begin{figure*}[t]
\centering
\includegraphics[width=10.0cm]{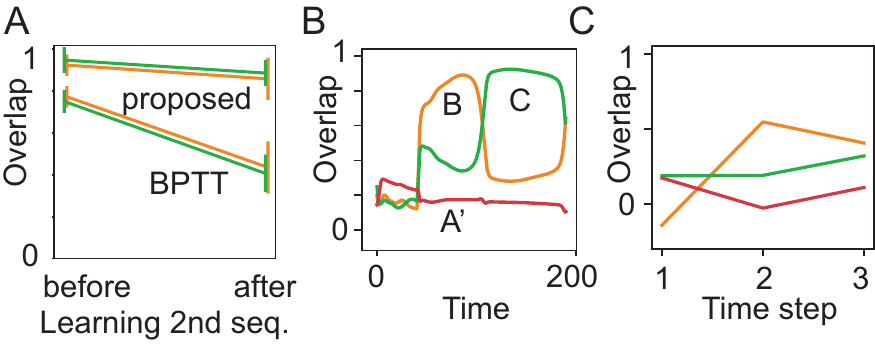}
\caption{
Comparison between behaviors in our proposed model and BPTT.
{\textbf A}: Overlaps with the targets $B$ and $C$ just before and after learning sequential patterns $(A',B)$ are plotted. 
Before-learning overlaps are measured in recalling $(A,B,C)$, while after-learning overlaps are in recalling $(A',B,C)$. 
Yellow and green lines represent the overlaps with $B$ and $C$, respectively.
Error bars are standard deviations for 10 trials of learning.
{\textbf B} and {\textbf C}: The neural dynamics in recall of  $(A',B,C)$ are plotted by using the overlaps after learning $A'$ and $B$ once for our model and in BPTT, respectively.
The overlaps of different targets are represented by different colors indicated in the panels.
}
\label{comp_BPTT}
\end{figure*}

We now consider the neural dynamics for $\boldsymbol{y}_{t=225}$, just after the transition from $\boldsymbol{\xi}^1_3$ to $\boldsymbol{\xi}^1_1$ (Fig \ref{fig:bif}C(i)).
For this $\boldsymbol{y}$, a single fixed point corresponding to the present pattern ($\boldsymbol{\xi}^1_1$) exists, leading to its stability against noise.
As $\boldsymbol{y}$ is changed, the basin of $\boldsymbol{\xi}^1_1$ shrinks, while a fixed point corresponding to the next target $\boldsymbol{\xi}^1_2$ appears, and its basin expands \footnote{The other fixed point corresponding to $\boldsymbol{\xi}^1_3$ also appears, but its basin is quite small. Thus, we can neglect this fixed point.}, as shown in Fig. \ref{fig:bif}C(ii).
At $\boldsymbol{y}_{t=290}$, the fixed point $\boldsymbol{\xi}^1_1$ becomes unstable.
Thus, the neural state $\boldsymbol{x}$ at $\boldsymbol{\xi}^1_1$ goes out of there, and falls on $\boldsymbol{\xi}^1_2$, i.e., a transition occurs.
If stronger noise is applied, this state will be kicked out of $\boldsymbol{\xi}^1_1$ earlier than in the noiseless case, resulting in a decrease in the duration of stay at the target (see Supplemental Material for details). 

With further shift of $\boldsymbol{y}$, $\boldsymbol{y}_{t=295,300,\cdots}$,
a regime of coexistence of $\boldsymbol{\xi}^1_2$ and $\boldsymbol{\xi}^1_3$ with large basins appears (Fig. \ref{fig:bif}C(iii)).
The basin of the attractor $\boldsymbol{\xi}^1_2$ shrinks and vanishes (Fig. \ref{fig:bif}C(iv)), and the transition from $\boldsymbol{\xi}^1_2$ to $\boldsymbol{\xi}^1_3$ occurs at $t=375$.
The next transition from $\boldsymbol{\xi}^1_3$ to $\boldsymbol{\xi}^1_1$ occurs in the same manner at $t=460$.
These processes provide the mechanism for robust sequential recall: 
fixed points $\boldsymbol{x}$ of the current and successive targets coexist, and then, the current target becomes unstable when the slow variables change.

\subsection*{Inference}
Next, we test if our model flexibly inferred new sequences based on the previously learned sequence.
To this end, we considered a simple task (See Materials and Methods for details). 
First, a network learns a sequence $(\boldsymbol{\xi^1_1},\boldsymbol{\xi^1_2},\boldsymbol{\xi^1_3})=(A,B,C)$.
The overlaps with targets reach more than 0.9 after 20 epochs of learning, as shown in Fig. \ref{comp_BPTT}A and Fig. S3A.
After learning the first sequence, the network learns a new sequence $(\boldsymbol{\xi^2_1},\boldsymbol{\xi^2_2})=(A',B)$.
If the network succeeds in using the already learned sub-sequences $B$ and $C$,
the sequence $(A',B,C)$ can be immediately generated.
The average overlap with the sequence $(A',B,C)$ is increased during learning, while that with the first sequence $(A,B,C)$ is slowly decreased (Figs. S3A).
Note that even after learning $A'$ and $B$ only once, the overlap with $C$ in the second sequence takes a high value.
As an example, we plot the fast dynamics of a network after learning $A'$ and $B$ once in Fig. \ref{comp_BPTT}B.
$A'$ evokes $B$ and $C$, although the overlap with the first target $A'$ is quite small.
Thus, our model is able to infer a new sequence based on the previously learned sequence. 

To elucidate the capacity of flexible inference in our model, we compared the performance with that by the back-propagation through time (BPTT) algorithm (See Material and Methods for details). 
BPTT gradually learns a new sequence by decreasing the total error over all targets.
Thus, we expected that the successive patterns $B$ and $C$ are generated on recalling the second sequence $(A',B,C)$.
Indeed, after learning $A'$ and $B$ once, the overlaps with $B$ and $C$ in the second sequence drastically decrease, meaning that BPTT does not use the previously learned sequence, as shown in Figs. \ref{comp_BPTT}A and C.
The overlaps with all of targets in the second sequence are quite small, as shown in Fig. \ref{comp_BPTT}C.
To sum, our model flexibly learns a new sequence by using the already learned structure, whereas BPTT does not.

\begin{figure*}[t]
\centering
\includegraphics[width=11.4cm]{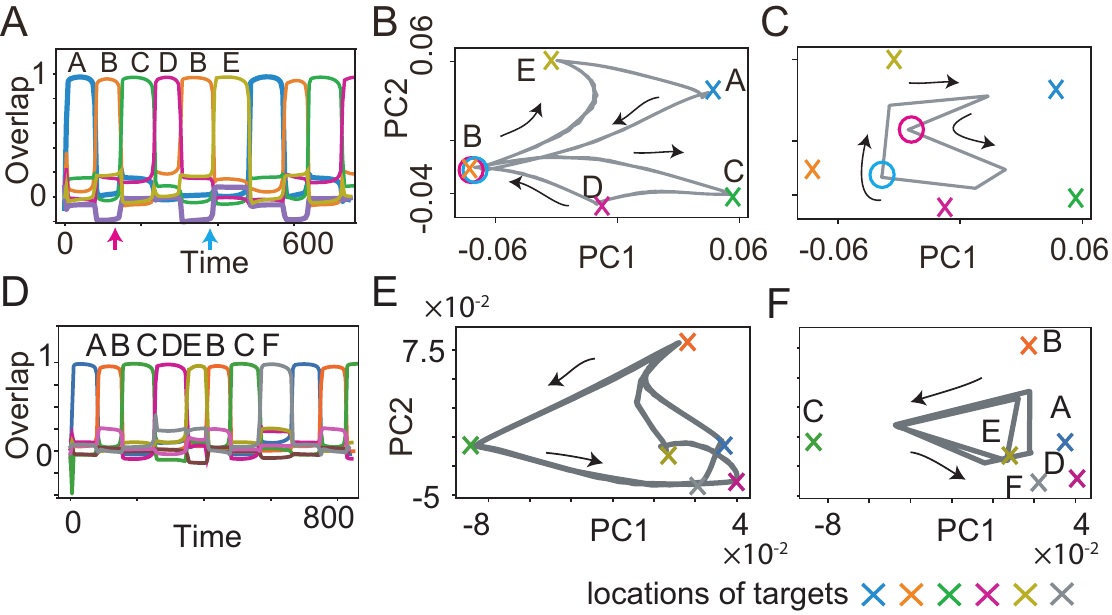}
\caption{
        Recall processes for history-dependent sequences for $K=1,M=6$ (in {\textbf A}-{\textbf C} and for $K=1,M=8$ (in {\textbf D}-{\textbf F}).
        {\textbf A} and {\textbf D}: The neural activities of $\boldsymbol{x}$ upon $\boldsymbol{\eta}^1$ are plotted by using their overlaps with the targets.
        Colors and alphabets indicate which of targets overlapped.
        {\textbf B} and {\textbf C} ({\textbf E} and  {\textbf F}): The neural dynamics plotted in A (D) are shown by projecting the fast dynamics in B (E) and the slow dynamics in C (F) onto a 2-dimensional PC space.
        X-shaped marks represent the locations of the targets.
        Magenta and cyan circles in B indicate the locations of $\boldsymbol{x}$, respectively, just before targets $C$ and $E$ are recalled (as indicated by the arrows in A), whereas the circles in C indicate the locations of $\boldsymbol{y}$.
        }
    \label{fig:comp-rcl}
\end{figure*}

\subsection*{Context-dependent learning}
We examined if the proposed model learns the history-dependent sequence ($M=6$),
in which the same patterns exist in a sequence such as 
$(\boldsymbol{\xi^1_1},\boldsymbol{\xi^1_2},\cdots,\boldsymbol{\xi^1_6})=(A,B,C,D,B,E)$.
The patterns succeeding $B$ are $C$ or $E$, depending on whether the previous pattern is $A$ or $D$.
Then, the neural dynamics have to retain the information of the target $A$ or $D$, to recall the target $C$ or $E$ correctly.
Our model succeeded in recalling this sequence, as shown in Fig. \ref{fig:comp-rcl}A.
Just before the target $C$ and $E$ are recalled, there is no clear difference in the values of fast variables $\boldsymbol{x}$, as indicated by the circles in Fig. \ref{fig:comp-rcl}B.
However, the values of slow variables $\boldsymbol{y}$ are different, depending on the previous targets shown in Fig. \ref{fig:comp-rcl}C, which stabilize different patterns of $\boldsymbol{x}$.
Furthermore, we demonstrate that our model succeeded in recalling more complex sequences ($M=8$) such as 
$(\boldsymbol{\xi^1_1},\boldsymbol{\xi^1_2},\cdots,\boldsymbol{\xi^1_8})=(A,B,C,D,E,B,C,F)$, as shown in Fig. \ref{fig:comp-rcl}D. 
In this case, the neural dynamics have to keep three previous targets in memory to recall the target $D$ or $F$ after $BC$ correctly.
Although the difference in the activities of $\boldsymbol{y}$ after recalling $ABC$ or $EBC$ is quite small,
it is sufficient to identify which of the sequences should be recalled.
By using this difference, the correct targets depending on the previous patterns $A$ or $E$ are stabilized.

\begin{figure}[t]
\centering
\includegraphics[width=8cm]{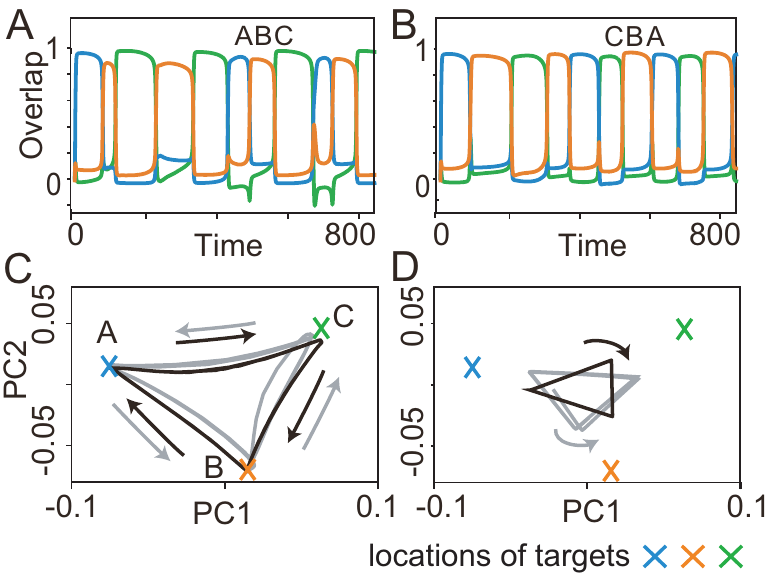}
\caption{
    Recall processes for history-dependent sequences for $K=2, M=3$. 
    {\textbf A}: The neural activities of $\boldsymbol{x}$ upon $\boldsymbol{\eta}^1$ (i) and $\boldsymbol{\eta}^2$ (ii) are plotted by using their overlaps with the targets.
    Colors and alphabets indicate the targets overlapped.
    {\textbf B} and  {\textbf C}: The neural dynamics shown in A are shown by projecting them onto a 2-dimensional PC space.
    The fast dynamics are shown in B and the slow dynamics are shown in C.
    The neural trajectories upon $\boldsymbol{\eta}^1$ and $\boldsymbol{\eta}^2$ are plotted in gray and black, respectively.
	    }
    \label{fig:comp-2ipt}
\end{figure}

As another example of the history-dependent sequences, we explored learning two history-dependent sequences (Fig. \ref{fig:comp-2ipt}), namely,
($\boldsymbol{\xi^1_1}, \boldsymbol{\xi^1_2}, \boldsymbol{\xi^1_3}$)=(A,B,C) upon $\boldsymbol{\eta}^1$, and
($\boldsymbol{\xi^2_1}, \boldsymbol{\xi^2_2}, \boldsymbol{\xi^2_3}$)=(C,B,A) upon $\boldsymbol{\eta}^2$.
In these sequences, the flow $A \rightarrow B \rightarrow C$ on the state space under $\boldsymbol{\eta}^1$ should be reversed under $\boldsymbol{\eta}^2$
\footnote{This task is not easy. The strength of the external input $\eta$ in Eq. \ref{eq:input} has to be tuned (around 1.3). The success rate is small, at just over $10\%$ }.
The learned network succeeds in generating these sequences.
Although orbits of $\boldsymbol{x}$ under different signals almost overlap in the 2-dimensional space, those of $\boldsymbol{y}$ do not.
This difference in $\boldsymbol{y}$, in addition to different context signals, allows the orbits of $\boldsymbol{x}$ in the reverse order of patterns.
Generally, $\boldsymbol{y}$ is different depending on the history of the previous patterns and inputs even when $\boldsymbol{x}$ is same. 
Different $\boldsymbol{y}$ stabilizes different fixed point of $\boldsymbol{x}$, to generate the history-dependent sequence.

\subsection*{Timescale dependence}
\begin{figure*}[t]
\centering
\includegraphics[width=11.4cm]{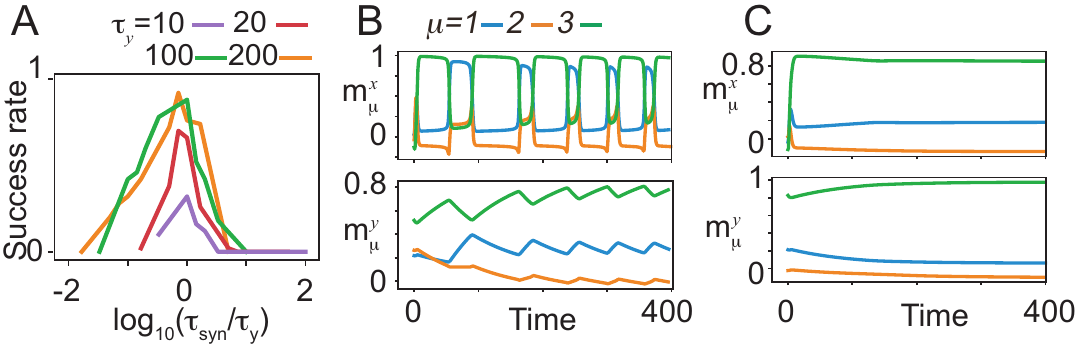}
\caption{
    Time scale dependence on neural dynamics.
    {\textbf A}: The success rate of recalls as functions of $\tau_{syn}$ for given $\tau_y$.
    The curves of the success rates are rescaled by $\tau_y$.
    Different colors represents different $\tau_y$ indicated by bars below panels.
    The success rate is calculated across fifty realizations for $K=1, M=7$.
    {\textbf B} and {\textbf C}: The neural activities of $\boldsymbol{x}$ (upper) and $\boldsymbol{y}$ (lower) in the recall process are shown by using the overlaps with the targets in the same color as shown in Fig. \ref{fig:bif}A.
    The neural dynamics for $\tau_y=100, \tau_{syn}=10$ are shown in B, while those for $\tau_y=100, \tau_{syn}=1000$ are shown in C.
}
    \label{fig:timescale}
\end{figure*}

Finally, we calculated the success rate of recalls as a function of $\tau_{syn}$ for different $\tau_y$ by fixing $\tau_x$ at 1, as are plotted after rescaling $\tau_{syn}$ by $\tau_{y}$ in Fig. \ref{fig:timescale}A.
The ratios yield a common curve that shows an optimal value $\sim 1$ at $\tau_{syn}$, approximately equal to $\tau_y$ \footnote{
For $\tau_{y} = 10$, which is close to $\tau_{x}$, the success rate yields a lower value for the optimal $\tau_{syn}$.}.
The balance between $\tau_{syn}$ and $\tau_{y}$ is important to regulate the success rate when they are sufficiently smaller than $\tau_{x}$.

To unveil the significance of the timescale balance, we, first, present how the recall is failed for $\tau_y >> \tau_{syn}$, ($\tau_y=100,\tau_{syn}=10$) in Fig. \ref{fig:timescale}B).
Some of the targets are recalled sequentially in a wrong order, whereas other targets do not appear in the recall process.
To uncover the underlying mechanism of the failed recall, we analyze the neural dynamics of fast variables with slow variables quenched in a manner similar to that shown in Fig. \ref{fig:bif} (See Fig. S\ref{fig:sup_timescale}).
Here, all the targets are stable for certain $\boldsymbol{y}$, although $\boldsymbol{\xi}^1_2$ does not appear in the recall process.
We also found that fixed points corresponding to $\boldsymbol{\xi}^1_1$ and $\boldsymbol{\xi}^1_2$ do not coexist:
the fixed point corresponding to $\boldsymbol{\xi}^1_3$ has a large basin across all $\boldsymbol{y}$.
This leads to a transition from $\boldsymbol{\xi}^1_1$ to $\boldsymbol{\xi}^1_3$ by skipping $\boldsymbol{\xi}^1_2$, and thus, the recall is failed.

Interestingly, failed recalls for $\tau_y << \tau_{syn}$ are distinct from those for $\tau_y >> \tau_{syn}$.
For $\tau_y=100,\tau_{syn}=1000$, only the most recently learned target is stable for almost all $\boldsymbol{y}$, and thus, only this target is recalled, as shown in Fig. S\ref{fig:sup_timescale}A.
We sampled the slow variables from the last learning step of the sequence (Fig. S\ref{fig:sup_timescale}B), and analyzed the bifurcation of the fast variables against change in slow variables, in the same way as above.
Here, only the latest target (here, $\boldsymbol{\xi}^1_3$) is a fixed point, whereas the other targets are not.
Thus, transitions between targets are missed, except the transition to the latest target.
These results indicate that the timescale balance changes bifurcation of the fast dynamics and the memory capacity.

\section*{Discussion}
Sequential transitions between metastable patterns are ubiquitously observed in the neural system \cite{Miller2016} during various tasks, such as
perception \cite{Jones2007,Miller2010}, decision making \cite{Ponce-Alvarez2012}, working memory \cite{Stokes2013,Taghia2018},and recall of long-term memory \cite{Wimmer2020}.
We have developed a novel neural network model with the fast and slow dynamics to generate sequences with non-Markov property and  concatenate sequences, which are based on these cognitive functions.
%The slow dynamics regulate stabilization and destabilization of the fixed points in the fast dynamics dependent on the stored preceding stimuli \cite{Kurikawa2013, Kurikawa2020}.

In a standard method for generating sequential patterns \cite{Kleinfeld1986, Sompolinsky1986, Nishimori1990, Seliger2003, Recanatesi2015, Russo2012, Haga2019}, 
asymmetric Hebbian learning between a pattern $\mu$ and the next $\mu+1$, i.e., $\boldsymbol{\xi}^{\mu+1} (\boldsymbol{\xi}^{\mu})^t$, is used to create the transition from $\boldsymbol{\xi}^{\mu}$ to $\boldsymbol{\xi}^{\mu+1}$ \cite{Kleinfeld1986, Sompolinsky1986, Recanatesi2015, Russo2012, Haga2019}.
In these studies, however, only the connections between the current and immediately preceding patterns are embedded in the connectivity, resulting in that the prolonged history of the patterns cannot be embedded. 
Thus, non-Markov sequences are not generated in contrast to our model\footnote{There are studies that finely designed networks to show the history-dependent sequences \cite{Verduzco-Flores2012, Chartier2006}.
However, they require the additional neurons or sub-networks as many as the number of memories or sequences.}.

In some models, a term that changes slower than the neural dynamics (e.g., an adaptation term) is introduced to lead to the transition.
In \cite{Gros2007, Recanatesi2015, Russo2012}, the slow term is introduced to destabilize the current pattern. However, this term does not determine the next pattern and,
thus, another mechanism is necessary for the transition to the desired pattern.
The feedback from the slow population in our model, in contrast, not only destabilizes the current pattern, 
but also simultaneously stabilizes the next targeted pattern.
As the current and next pattern coexist for some time span, the robust transition between them is achieved.

Alternatively, supervised learning methods used in machine learning fields, such as BPTT \cite{Werbos1990}, are investigated to reproduce sequential neural activities observed experimentally \cite{Mante2013,Carnevale2015,Chaisangmongkon2017}, including non-Markov trajectories \cite{Sussillo2009,Laje2013}.
BPTT learning, however, cannot concatenate the previously learned sub-sequence to the newly learned one, 
due to the catastrophe forgetting.
Further, the BPTT requires non-local information, which is biologically implausible, whereas the trajectories shaped by this method are vulnerable to noise \cite{Laje2013}.
Our model is free from these deficiencies.

Timescales in the neural activities are hierarchically distributed across several cortical areas  \cite{Hasson2015, Honey2012, Murray2014, Runyan2017}. 
For instance, consider the hippocampus (HPC) and the prefrontal cortex (PFC), which are coupled by mono-synaptic and di-synaptic connections \cite{Ito2015}.
HPC neurons respond to the location of animals \cite{Kumaran2016} with faster timescales than those in PFC, which has the slowest timescale among cortical areas \cite{Murray2014}.
Experimental studies \cite{Guise2017,Ito2015} revealed that PFC neurons are necessary to differentiate HPC dynamics depending on the context and previous experience.  
Similarly, neurons in the orbitofrontal cortex (OFC), whose timescales are considered to be slower than those in HPC, are necessary for concatenating the sequences in the stimulus-reward response \cite{Jones2012,Wikenheiser2016}.
Accordingly, it is suggested that the area with the slow dynamics is necessary to generate and concatenate the sequences.

%We have demonstrated a novel method for generating the sequences:
%the slow dynamics stabilizes and destabilizes patterns in the fast dynamics.
%In our model, the neural state space in the slow dynamics is separated into different regimes in which different patterns in the fast dynamics are stabilized.
%We predict that the neural activities in the slow dynamics area such as PFC and OFC show clear boundaries, at which the neural activities in the fast dynamics area such as HPC changes their patterns drastically.

Neural networks with multiple timescales are investigated theoretically in several studies.
In some studies \cite{Perdikis2011, Yamashita2008}, the slow dynamics are introduced to concatenate primitive movements and produce a complex movement, while hidden states of the hierarchical external stimuli are inferred by the multiple timescales in the neural dynamics in another study \cite{Kiebel2009}.
In \cite{Perdikis2011,Kiebel2009}, the relationship between the slow and fast dynamics are fixed a priori to perform their tasks, whereas, in our model, such a relationship is shaped through the learning process.
In \cite{Yamashita2008}, the BPTT method is adopted for training the network; thus it faced the same drawbacks as already mentioned.

%As for the timescales, we need further studies to fill a gap between our model and experimental observations. 
%We found the optimal relationship between the timescales of the fast, slow neural dynamics and the synaptic plasticity.
%For the optimal relationship, the timescale in the slow dynamics are at least 20 times slower than that in the fast and comparable with that of the synaptic plasticity.
%Electro-physiological studies\cite{Wang2016} revealed that the timescale in neural activity in the prefrontal cortex is on the order of (less than) a second,
%while that in sensory cortex is on the order of (larger than) hundreds of milliseconds. 
%The ratio of the timescale in the slow dynamics to that in the fast is less than 10 times.
%Further, typical synaptic plasticity, long term potentiation\cite{Bliss1973} are induced by strong stimulation in a minute at best\cite{Bayazitov2007}.

As for the timescales, we need further studies to fill a gap between our model and experimental observations. 
The ratio of the timescale in the slow dynamics to that in the fast dynamics is less than 10 times across cortical areas \cite{Wang2016}, which are smaller than the optimal ratio in our model.
Further, the difference between the timescales in the slow dynamics (on the order of a second) and in the synaptic plasticity (on the order of a minute \cite{Bliss1973,Bayazitov2007}) is larger than that adopted in our model.

Diversity in the timescales of individual neurons and the calcium dynamics possibly resolve this discrepancy. 
The timescale of individual neurons in the same area is distributed over two digits \cite{Wasmuht2018,Bernacchia2011}.
%dominantly contribute to generating sequential patterns, the required relation in the timescale between the fast and slow neurons is in line with our model.
The calcium dynamics in the synapses can modify the synaptic efficacy on the order of a second \cite{Graupner2012,Shouval2010}.
By taking these effects into account, our model may be consistent with the experimental observations,
although further studies will be important, including those with spiking neurons \cite{Kurikawa2015} and spike-timing-dependent potentiation. 

%\textcolor{red}{Regarding to the synaptic plasticity, theoretical studies\cite{Graupner2012,Shouval2010} indicated the calcium dynamics in the pre- and postsynaptic neurons are driven by neural activity and, in turn, control the synaptic plasticity on the order of a second.}

%and presents a basic general framework of how a multiple-timescale system provides history-dependent sequential patterns, essential to cognitive functions.

\section*{Materials and Methods}
\subsection*{Details of learning procedure in the inference task}
Here, in contrast to other tasks, we presented sequentially different inputs for a sequence.
For the first sequence $(A,B,C)$, we applied $\eta^1_1 = a$ for target $A$, and $\eta^1_2 =\eta^1_3=b$ for the targets $B$ and $C$.
For the second sequence $(A',B,C)$, in the same manner, we applied $\eta^2_1 = a'$ for the target $A'$, and $\eta^2_2 =\eta^2_3=b$ for the targets $B$ and $C$.
All the targets and inputs are randomly sampled according to the probability $P[ (\xi^{\alpha}_{\mu})_{i}=\pm 1]=P[ (\eta^{\alpha}_{\mu})_{i}=\pm 1]=1/2$.
Further, we modified the procedure in the exchange timing of the inputs and targets in the learning process.
We changed the inputs and targets at $T=40$ after beginning of learning the first patterns in both sequences.

For reference, we also used backpropagation through time (BPTT) algorithm with Adam to train a recurrent network.
We build a three-layer recurrent network model including input ($N_{in}$ neurons), hidden ($N_{hid}$ neurons), and output layers ($N_{out}$ neurons).
Input-hidden $W_I$, hidden-hidden (recurrent) $W_r$, hidden-output $W_O$ connections are all-to-all connections and modified by BPTT.
$N_{in} \times N_{hid} + N_{hid} \times N_{hid} + N_{hid} \times N_{out}$ is the number of parameters to be tuned.
We set $N_{in}=N_{out}=100$ and $N_{hid}=40$ to match the number of tuned parameters in BPTT as that in our model.
The activity of each element is updated according to the following equation:
\begin{align*}
  y_{i,t} &= \tanh{(\Sigma_j (W_I)_{ij}I_{j,t} + \Sigma_j (W_r)_{ij} y_{j,t-1})}, \\
  z_{i,t} &= \tanh{( \Sigma_j (W_O)_{ij} y_{j,t})},
\end{align*}
where $I_{i,t}$, $y_{i,t}$ and $z_{i,t}$ are the activities of $i$-th elements in the input pattern, the hidden and output layer at time $t$, respectively.
The loss function is $L=\Sigma_{i,t} (\xi_{i,t} - z_{i,t})^2/2$.
Here, $\xi_{i,t}$ is the value of $i$-th element of the target at time $t$.
Inputs and targets are same to those in our model;
$(I_{1},I_2,I_3) = (a,b,c)$ and $(a',b,c)$ for the fast and the second sequence, respectively, while $(\boldsymbol{\xi}_1,\boldsymbol{\xi}_2,\boldsymbol{\xi}_3) = (A,B,C)$ and $(A',B,C)$ for the fast and the second sequence, respectively.
The learning parameter in BPTT is set to $0.001$ and the discount rates of the first and second moments in Adam are set to $0.99$ and $0.999$, respectively.

\section*{Acknowledgments}
We thank Omri Barak and Rei Akaishi for fruitful discussion and Tatsuya Haga for useful comments on our manuscript.
This work was partly support by JSPS KAKENHI (nos. 18K15343 and 20H00123).

%\clearpage

%This is where your bibliography is generated. Make sure that your .bib file is actually called library.bib
\bibliography{10th_seq.bib}

%This defines the bibliographies style. Search online for a list of available styles.
\bibliographystyle{abbrv}

\newpage 
\renewcommand{\figurename}{Fig.S}
\setcounter{figure}{0}
\section{Supplemental text}

\subsection{Learning multiple sequences}
Our model memorizes several sequences for different context signals.
We exemplify the procedure of memorization by focusing on the learning and recall process for $K=2$.
Learning two sequences are accomplished in a manner similar to the learning of a single sequence, as described in the main text.
The model learns two sequences alternatively: the first sequence ($\boldsymbol{\xi}^1_1,\boldsymbol{\xi}^1_1,\cdots,\boldsymbol{\xi}^1_{M},\boldsymbol{\xi}^1_1$) is learned with the same criteria for $K=1$.
After resetting the fast and slow variables, the second sequence ($\boldsymbol{\xi}^2_1,\boldsymbol{\xi}^2_2,\cdots,\boldsymbol{\xi}^2_{M},\boldsymbol{\xi}^2_1$) is learned in the same way. 
We repeated these processes 20 times before finishing the learning.

\begin{figure*}[thp]
  \begin{center}
     \includegraphics[width=150mm]{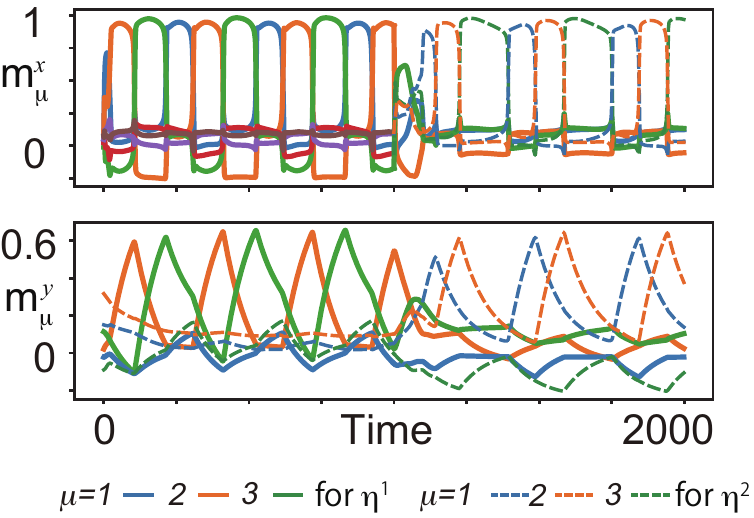}
    	\caption{
    	   Recall dynamics for $K=2, M=3$ when the context is switched at $t=1000$.
    	   The fast variables (top) and slow variables (bottom) are plotted by using the overlap $m^{\alpha}_{\mu}$ ($\alpha=1,2$ and $\mu=1,2,3$).
    	   The index of the overlap is indicated below the panels.
    	   }
  \label{fig:sup_rcl}
  \end{center}
  \end{figure*}

Fig. S\ref{fig:sup_rcl} shows a recall process for $K=2,M=3$ after learning.
In the presence of $\boldsymbol{\eta^{1}}$, the sequence ($\boldsymbol{\xi^1_1},\boldsymbol{\xi^1_2},\boldsymbol{\xi^1_3}$) is recalled, as shown in the figure.
Then, after switching the input from $\boldsymbol{\eta^{1}}$ to $\boldsymbol{\eta^2}$ at $t=1000$,
the required sequence ($\boldsymbol{\xi^2_1},\boldsymbol{\xi^2_2}, \boldsymbol{\xi^2_3}$) is recalled successfully.

\subsection{Robustness of the sequences}
\begin{figure*}[htp]
  \begin{center}
     \includegraphics[width=150mm]{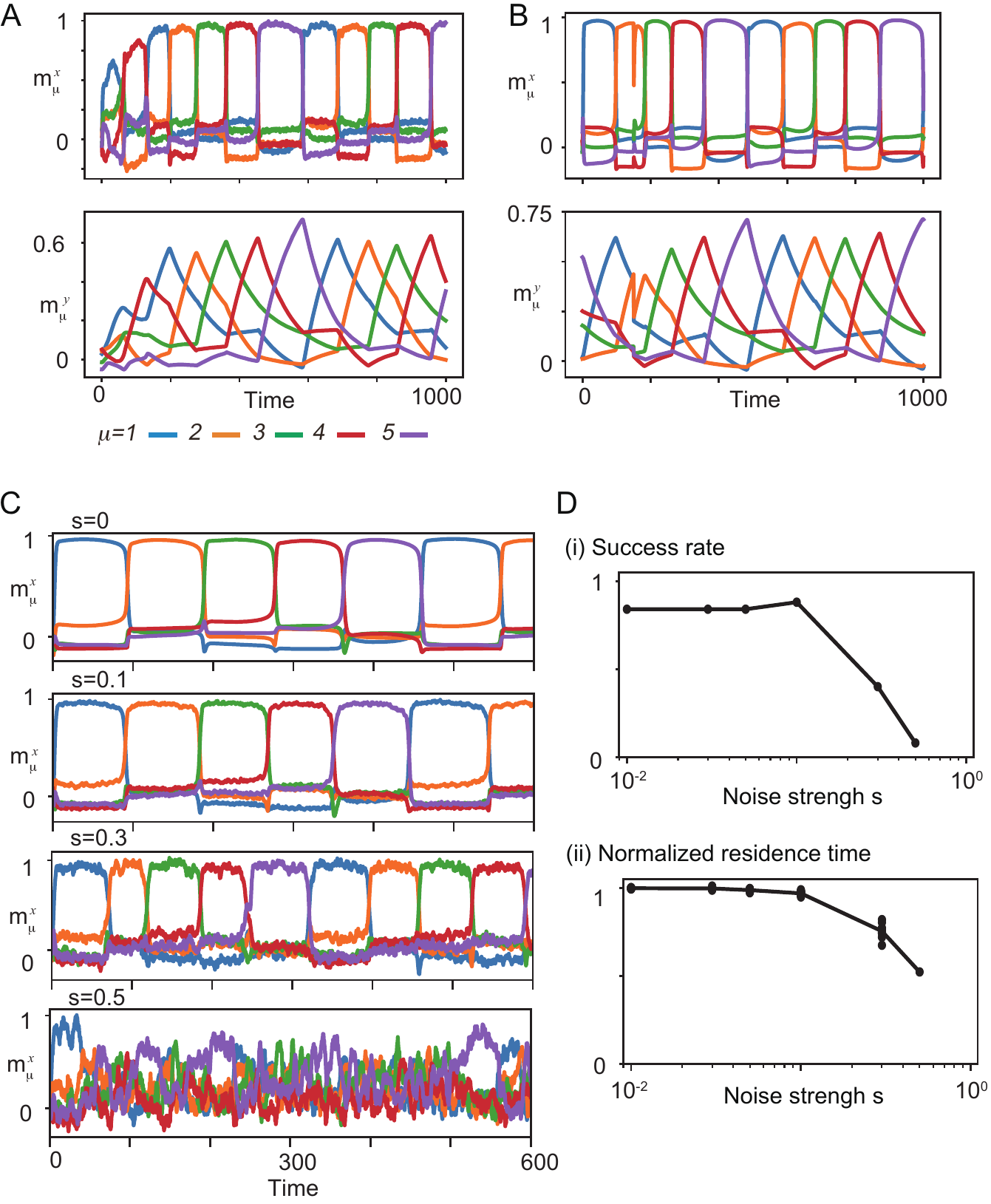}
    	\caption{
    	  {\textbf A} and {\textbf B}: Trajectory of the overlaps of $\boldsymbol{x}$ (upper panel) and $\boldsymbol{y}$ (lower panel).
    	  The trajectory with the targets for noise strength $s=0.1$ is shown in A, and that with one-shot perturbation (at $t=150$) is shown in B. 
    	  Each color indicates the target used in the calculation of the overlap (the same color code is used in the following panels).
    	  {\textbf C}: The time series of $\boldsymbol{x}$ are plotted by using the overlaps. 
    	  The realization of the network, the target, and the context signal patterns are identical across panels, whereas the noise strength $s$ is increased from the upper to the lower panels.
    	  {\textbf D}: The success rate and the normalized residence time at each pattern is plotted against the noise strength $s$ in (i) and (ii), respectively.
    	  The success rate is defined in the same manner as in Fig \ref{fig:timescale}A, and calculated across twenty-five realizations of networks, targets, and context patterns.
    	  The normalized residence time is defined in the text, and obtained only from the successful recalls in the twenty-five realizations.
    	  The dots in (ii) indicate the durations of different realizations.
	}
  \label{fig:sup_noise}
  \end{center}
  \end{figure*}

We investigated the robustness in the sequence recall.
First, we applied strong one-shot perturbations into the neural dynamics,
where multiplicative noise was added to the neural activities of all neurons ${x_i}$ and ${y_i}$, as $x_i \rightarrow (1-r_i^x)x_i$, $y_i \rightarrow (1-r_i^y)y_i$ ($i=1,2,\ldots,N$), and $r^{x,y}_i$ was chosen randomly from a uniform distribution of 0 to 1.
The trajectory with the one-shot perturbation is shown in Fig. S\ref{fig:sup_noise}B.
After the perturbation, the neural dynamics rapidly recover to a limit cycle, in which the neural activity exhibits transition from one target to another in the correct order.

Next, we examined the robustness against the change in the initial states and noise. 
Here, Gaussian white noise $\boldsymbol{\zeta}(t)$ was added into the neural dynamics $\boldsymbol{x}$ and $\boldsymbol{y}$ given by Eqs. (\ref{eq:neuro-dyn},\ref{eq:neuro-dyn-slow}) with satisfying $<\zeta_i(t) \zeta_j(t')>=s \delta_{ij} \delta(t-t')$ for $i=j$; otherwise $0$.
Here, $\delta_{ij}$ and $\delta$ are the Kronecker and Dirac delta, respectively, and $s$ is the noise strength.
Fig. S\ref{fig:sup_noise}A shows a trajectory from a random initial state under the noise by using the overlaps of the slow and fast variables for $K=1,M=5$.
After the transient period, the trajectory converges to the limit cycle that generates the correct sequence recall.
We tested nine other trajectories under noise from nine random conditions, and found that all trajectories converge to the limit cycle.

Furthermore, the robustness of the model against noise strength was examined.
The dynamics of $\boldsymbol{x}$ for increasing the noise strength are plotted in Fig. S\ref{fig:sup_noise}C.
Below $s=0.3$, the sequence is recalled with the correct order.
For stronger noise ($s=0.5$), only a few patterns are recalled intermittently, and others are not.
Fig. S\ref{fig:sup_noise}D(i) shows the success rate of recalls as a function of the noise strength. The success rate is approximately 0.8 (same as the ratio in the case without noise) up to $s=0.1$, and decreases rapidly.
All of these results demonstrate that the sequential patterns in our model are quite robust against changes in the initial states and noise.

Finally, we measured the duration for which the fast dynamics stay on each target.
The duration measured with noise is normalized by that measured without noise.
The normalized duration is plotted as a function of the noise strength in Fig. S\ref{fig:sup_noise}D(ii).
We found that the normalized duration decreases as the noise strength increases.
After the fast dynamics converge to the target attractor, the basin volume reduces over time, as shown in Fig. \ref{fig:bif}.
Thus, stronger noise is likely to kick out the neural states from the targets earlier, resulting in a decrease in duration as the noise strength increases. 

\section{Supplemental figures}

\begin{figure*}[htp]
  \begin{center}
     \includegraphics[width=150mm]{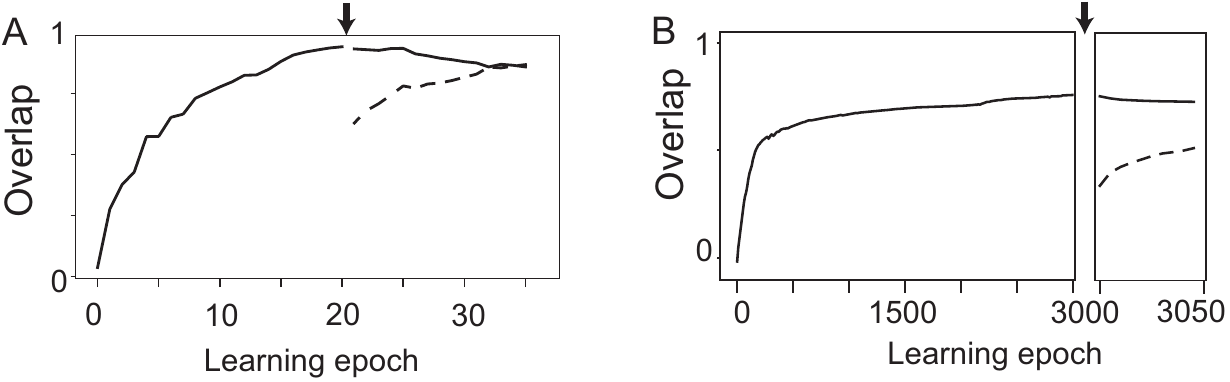}
    	\caption{
    	  Learning performance of two sequences for our learning model ({\textbf A}) and BPTT ({\textbf B}).
    	  The learning performance is measured by average value of the overlaps of the fast dynamics with sequence patterns.
The filled and broken lines represent overlaps with the first $(A,B,C)$ and second sequences $(A',B,C)$, respectively.
The arrows on the panel show beginning of learning the sequential patterns $(A',B)$.    }
  \label{fig:sup_comp}
  \end{center}
  \end{figure*}

\begin{figure*}[htp]
  \begin{center}
     \includegraphics[width=150mm]{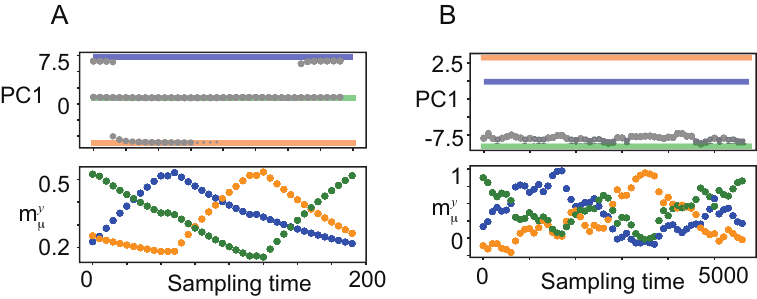}
    	\caption{
    	  {\textbf A} and {\textbf B}: The bifurcation diagrams of the fast variables with quenched $\boldsymbol{y}$ are shown for $(\tau_y,\tau_{syn}) = (100,10)$ and $(100,1000)$, respectively.
    	  These diagrams are plotted in basically same manner, as shown in Fig. \ref{fig:bif}B.
    	  At a different point from the analysis in Fig. \ref{fig:bif}B for $\tau_y=100,\tau_{syn}=100$, the slow variables are sampled from the trajectory in the final learning step of the sequence (namely, after learning the sequence nineteen times), because all targets do not appear in the recall process.
       	  The fixed points are plotted as circles, and colored lines represent the locations of the target 1 (blue), 2 (orange), and 3 (green) by projection onto the 1st principle components in the upper panels by the principle component analysis.
    	  In the lower panels, sampled $\boldsymbol{y}$ from the learning process are plotted by using the overlaps with the same color codes as in Fig. \ref{fig:timescale}B.
    }
  \label{fig:sup_timescale}
  \end{center}
  \end{figure*}

\end{document}